\documentclass[preprint,12pt,groupaddress]{elsarticle}

\usepackage{booktabs}
\usepackage{color}
\usepackage{lipsum}
\usepackage{float}
\usepackage[titletoc,toc,title]{appendix}
\usepackage{amssymb}
\usepackage{epsfig}
\usepackage{amsmath}
\usepackage{subfigure}
\usepackage{graphicx}
\usepackage{booktabs}
\usepackage{colortbl}
\usepackage{natbib}
\usepackage{color}
\usepackage{bbm}

\makeatletter
\def\ps@pprintTitle{%
 \let\@oddhead\@empty
 \let\@evenhead\@empty
 \def\@oddfoot{}%
 \let\@evenfoot\@oddfoot}
\makeatother

\begin{document}

\title{The thermodynamic efficiency of the Lorenz system}

\author[label1]{\'{A}lvaro G. L\'{o}pez}
\author[label2]{Fernando Benito}
\author[label3]{Juan Sabuco}
\author[label5]{Alfonso Delgado-Bonal}

\address[label1]{Nonlinear Dynamics, Chaos and Complex Systems Group.\\Departamento de F\'isica, Universidad Rey Juan Carlos, Tulip\'an s/n, 28933 M\'ostoles, Madrid, Spain}

\address[label2]{Gentleman Scientist, Calle del Puerto de Somosierra 58, 28770 Colmenar Viejo, Madrid, Spain}

\address[label3]{School of Geography and the Environment, University of Oxford, Oxford OX1 3QY, UK}

\address[label4]{Smith School of Enterprise and the Environment, University of Oxford, Oxford OX1 3QY, UK}

\address[label5]{Universities Space Research Association, Columbia, MD, United States}

\date{\today}

\begin{abstract}
We study the thermodynamic efficiency of the Malkus-Lorenz waterwheel. For this purpose, we derive an exact analytical formula that describes the efficiency of this dissipative structure as a function of the phase space variables and the constant parameters of the dynamical system. We show that, generally, as the machine is progressively driven far from thermodynamic equilibrium by increasing its uptake of matter from the environment, it also tends to increase its efficiency. However, sudden drops in the efficiency are found at critical bifurcation points leading to chaotic dynamics. We relate these discontinuous crises in the efficiency to a reduction of the attractor's average value projected along the phase space dimensions that contribute to the rate of entropy generation in the system. In this manner, we provide a thermodynamic criterion that, presumably, governs the evolution of far-from-equilibrium dissipative systems towards their self-assembly and synchronization into increasingly complex networks and structures.\\
\end{abstract}

\maketitle
\renewcommand{\bottomfraction}{.99}

\newcommand{\nobracket}{}
\newcommand{\nocomma}{}



\section{Introduction}

The concept of \emph{dissipative structure} was coined by the physical chemist Ilya Prigogine and his collaborators in the late 1960s to designate \emph{open} physical systems that operate far from thermodynamic equilibrium as a consequence of the continuous flow of matter and energy from and towards their environment \cite{pri78}. Among these exchanges, open systems dissipate energy in terms of irreversible heat flow to their surroundings, a process that is always accompanied by the generation of entropy. Dissipative structures are ubiquitous throughout all the scales of the physical realm, from fundamental electrodynamic particles \cite{lop20,lop21} to astrophysical systems, such as Cepheid variable stars \cite{zhev63} and, more generally, stars and black holes \cite{car14,mah16}. We can find them in autocatalytic reactions and chemical clocks \cite{bel58,zha64}, in thermoacoustic oscillators \cite{mat03}, in turbulent fluids where vortexes and convective phenomena occur \cite{ben19,ray16}, or in the physics of lasers \cite{hak75}, for example. We can also find them in the domain of life sciences, as in biogeochemical cycles \cite{fal08}, in the metabolism of cells \cite{sel68}, in the cell cycle \cite{fer11}, in neuronal dynamics \cite{fit61}, or in reaction-diffusion systems \cite{tur90}. Finally, the business cycles \cite{kal35} in human societies and the more general ecological cycles \cite{vol26} can be classified as non-conservative systems as well.

All these nonlinear dynamical systems can be pushed away from their homogeneous equilibrium states by modifying the intake and release rates of energy and matter from the surroundings or, alternatively, by switching between various modes of dissipation. From a mathematical point of view, these shifts correspond to variations in the values of the model key parameters. As a consequence, these systems typically experience universal dynamic bifurcations that trigger the instability of their equilibrium states, prompting nonlinear oscillations. This transition to instability frequently involves \emph{symmetry breaking} \cite{and72}, giving rise to the appearance of stable coherent dynamical structures, which emerge from the chaotic fluctuations that take place at the underlying microscopic scale. In turn, with their coherent motion, these structures reduce the energy gradients enforced by the environment, converting \emph{exergy} to work and relaxing such thermodynamic stresses \cite{sch94}. 

When dissipative systems are pushed further from equilibrium, typical chaotic behavior frequently reemerges at the macroscopic scale. Then, chains of similar systems might couple through diffusive currents. This coupling can lead again to long-range correlations, coherent behavior, and even dynamical synchronization at higher levels of organizational hierarchy. To some extent, this fact would explain the bottom-up self-similar organization of nature. Furthermore, some authors have proposed that the ability of open physical systems to better compensate gradients and degrade exergy as instability increases is consubstantial to them \cite{sch94}. This is very important in ecological systems since they can be regarded as enormous dissipative structures integrated by many highly interconnected elements, which self-organize as a whole to cool down the energy gradient imposed by the sun \cite{sch94}. 

Despite their ubiquity, the study of self-excited systems as thermodynamic engines has been barely emphasized \cite{cor36,jen13}, being  mostly side-tracked in the modern literature. This is perhaps due to the fact that these nonlinear dynamical systems are many times treated in a rather mathematical and abstract fashion, without paying due attention to their detailed physical significance. However, very recent works have conceptually unified all these non-equilibrium dynamical systems under the principle of \emph{local activity} \cite{mai13} on the one hand, and the concept of \emph{self-oscillation} on the other \cite{jen13}. Although there are several bifurcation routes that can lead to self-organized phenomena in non-equilibrium systems, two outstanding ones are the Hopf and the Pitchfork bifurcations. Both universal routes can lead to oscillatory and rotatory motion and, in extended systems, to the development of autowaves, the formation of solitons, and spatiotemporal chaos \cite{mai13}. 

In the present work we use a mathematical model representing a real thermodynamic machine called the Malkus-Lorenz wheel \cite{mis06,ill12,kim17,kar19} to study its \emph{efficiency} and rate of \emph{entropy generation} as it is driven away from equilibrium. The driving is achieved by affecting the intake and release of matter from and to the environment. This machine was developed by the engineers Willem Malkus and Louis Howard at the MIT during the 1970s. Mathematically, it is equivalent to the paradigmatic Lorenz system introduced by Edward Lorenz to unveil the complex hidden order that is inherent to chaotic dynamics \cite{lor63}. As it is well-known, this system is of historical importance because it is intimately connected to the problem of atmospheric convection. Since the problem of Rayleigh-B\'enard convection is also of paramount relevance in the study of \emph{far-from-equilibrium} dynamical systems \cite{ben19}, constituting an archetypal thermodynamic structure, we expect that our results shall have widespread applicability to most, if not all, open physical dissipative systems. 
 
\section{The Malkus-Lorenz wheel}

\subsection{Model's description}

\begin{figure}
\centering
\includegraphics[width=1.0\linewidth,height=0.4\linewidth]{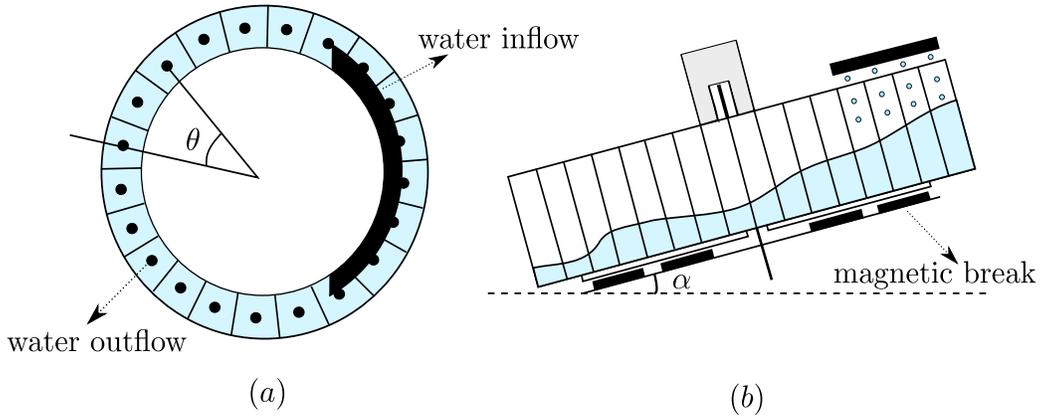}
\caption{\textbf{The Malkus-Lorenz wheel}. (a) Overhead view of the wheel, showing a sequence of compartments with holes at their bottom (black dots). A black hose injecting water is shown on the right, and the angle of rotation $\theta$ has been specified for clarity. (b) Lateral view of the wheel, with the tilting angle $\alpha$.}
\label{fig:1}
\end{figure}

The Malkus-Lorenz wheel consists of a series of compartments placed along a ring attached to a rotating wheel \cite{stro15}. As shown in Fig.~\ref{fig:1}, the wheel is tilted at an angle $\alpha$ from the floor. The position of each compartment can be referred to by its angular position $\theta$ along the ring. The ones at the top of the wheel are continuously fed by a flow of water at a constant rate $Q(\theta)$, entering from the nozzles of a perforated hose located above them. Each compartment presents a comparatively small hole at the bottom, through which water is excreted permanently. Thin tubes are attached to these holes so that the outward flow is essentially Poiseuille-like. In this manner, the mass leakage rate can be approximated as proportional to the angular mass density $\mu(\theta)$ of each compartment, with $K$ being the fractional leakage rate. An adjustable magnetic brake allows controlling the rate of energy dissipation of the wheel. This loss of energy is represented using a linear model, in which the frictional torque is proportional ($\nu$) to the velocity of the wheel, as suggested by Faraday's law. Finally, the gravitational field $g$ produces the necessary torque that makes the waterwheel rotate with a certain angular velocity $\omega(t)$.

As it is evident, this thermodynamic machine maintains its cyclic motion by taking water with high potential energy from the top and filtering water at lower potential energy through the holes at the bottom of the compartments. The energy dissipated due to friction from the magnetic brake flows into the surroundings in the form of heat. If the compartments are sufficiently small along the ring, we can represent the dynamics of the fluid by using the continuity equation for mass conservation
\begin{equation}
\dfrac{\partial \mu(\theta,t)}{\partial t}=Q(\theta)-K \mu(\theta,t)-\dfrac{\partial}{\partial \theta}\mu(\theta,t) \omega(t),
\label{eq:1}
\end{equation}
together with Euler's second equation describing the rotation of the wheel, which yields the integro-differential equation
\begin{equation}
I \dot{\omega}(t)=-\nu \omega(t)+g r \sin\alpha \int_0^{2 \pi} \mu(\theta)\sin\theta d\theta,
\label{eq:2}
\end{equation}
where we have assumed that the moment of inertia of the wheel is fixed, what holds once the transient dynamics has exhausted. If we bear in mind that the system is rotationally periodic, we can express its solutions in terms of the following Fourier series expansion
\begin{equation}
\mu(\theta,t)=\sum_{n=0}^\infty(a_n(t)\sin(n\theta)+b_n(t)\cos(n\theta)).
\label{eq:3}
\end{equation}
Assuming that the inflow of water is symmetrical, we can expand the function $Q(\theta)$ as an even Fourier series with coefficients $q_n$. The resulting set of equations can be written for all the modes of the wheel as
\begin{equation}
    \begin{array}{lr}
       \dot{a}_n(t)= n \omega(t) b_n(t)-K a_n(t) \bigskip\\
       \dot{b}_n(t)= -n \omega(t) a_n(t)-K b_n(t) + q_n  \bigskip\\
       \dot{\omega}(t)=-\dfrac{\nu}{I} \omega(t)+\dfrac{\pi g r \sin\alpha}{I} a_1(t)
     \end{array}.
     \label{eq:4}
\end{equation}
Note that, because the torque produced by gravity changes harmonically along the ring of the wheel, only the first order terms of the Fourier expansion of the mass distribution couple to the rotational motion. This yields a three-dimensional ODE system comprised by the equations
\begin{equation}
    \begin{array}{lr}
       \dot{a}_1(t)=  \omega(t) b_1(t)-K a_1(t)   \bigskip\\
       \dot{b}_1(t)= -\omega(t) a_1(t)-K b_1(t) + q_1  \bigskip\\
       \dot{\omega}(t)=-\dfrac{\nu}{I} \omega(t)+\dfrac{\pi g r \sin\alpha}{I} a_1(t)
     \end{array}.
     \label{eq:5}
\end{equation}
It can be mathematically demonstrated that this system is related to the Lorenz equations after an affine transformation, yielding the system of differential equations
\begin{equation}
    \begin{array}{lr}
       \dot{x}(t)= \sigma(y(t)-x(t)) \bigskip\\
       \dot{y}(t)= \rho x(t)-x(t)z(t)-y(t) \bigskip\\
       \dot{z}(t)= x(t)y(t)-z(t) 
     \end{array},
          \label{eq:6}
\end{equation}
where the variable $x(t)$ is related to the angular velocity of rotation, the variable $y(t)$  relates to $a_1(t)$, and $z(t)$ can be expressed in terms of $b_1(t)$. Furthermore, the nondimensional Rayleigh $\rho=\pi g r q_1/\nu K^2$ and Prandtl $\sigma=\nu/I K$ numbers have been introduced. We highlight the importance of Rayleigh number, since the thermodynamic efficiency of the water wheel can be expressed as a function of this parameter alone, when it is far from equilibrium in the regular regime, before the onset of chaos. It shall be used to represent the most remarkable bifurcation diagrams of the present work.
\begin{figure}
\centering
\includegraphics[width=0.5\linewidth,height=0.44\linewidth]{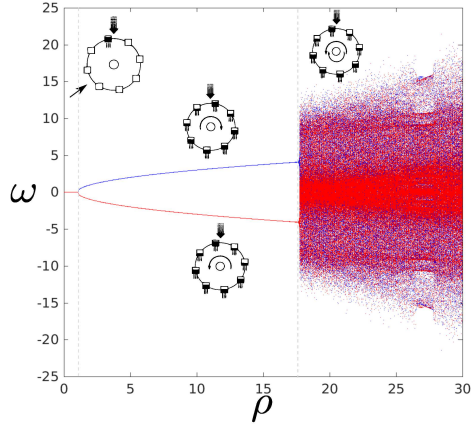}
\caption{\textbf{Bifurcation diagram}. The values of $\omega(t)$ at the various limit sets of the system as the Rayleigh parameter $\rho$ is varied. We distinguish three regimes. Firstly, a wheel at rest for $\rho<1$ can be seen. Secondly, a supercritical Pitchfork bifurcation befalls at $\rho=1$, and we find a wheel spinning uniformly clockwise or anticlockwise, depending on the arm of the diagram selected. Finally, a subcritical Hopf bifurcation takes place for $\rho_c=17.5$. The motion of the wheel becomes chaotic, alternating unpredictably from a clockwise to an anticlockwise rotation.}
\label{fig:2}
\end{figure}

\subsection{Model dynamical behaviour}

To better capture the effects of the dynamics of the Malkus-Lorenz waterwheel on its thermodynamic efficiency, we briefly describe the most significant features of the system of Eqs.~\eqref{eq:5}. To this end, we represent the bifurcation diagram of the angular velocity of the wheel as the Rayleigh number increases (see Fig.~\ref{fig:2}). This diagram describes the evolution of the fixed points as we drive the system far from equilibrium. From a physical point of view, this is achieved by progressively increasing the income of water by raising the value of $q_1$. This effect can be equally achieved by augmenting the radius of the wheel $r$, which leaves unmodified the value of the Prandtl number. In this manner, we do not have to worry about the fact that, at some point, we might overfeed the wheel, with the corresponding spill of the water. We can distinguish three fundamental dynamical regimes:

\begin{enumerate}

\item As can be seen in Fig.~\ref{fig:2}, as long as $\rho<1$, there exists only one stable fix point with values $\omega^{*}=0$, $a_1^{*}=0$ and $b_1^{*}=q_1/K$, which corresponds to a waterwheel at rest. In this regime, the water is evacuated too fast in comparison to the rate at which it is fed. No room is left for the development of instabilities. Any perturbation of the angular velocity is eventually eradicated, leading to a global attractor. This attractor corresponds to a passive system in a dead state of equilibrium. Importantly, we highlight that this state is not an equilibrium state but a stationary one from a thermodynamic point of view. A gradient imposed on the system by the water inflow is conducted through the compartments. Similarly, in the B\'enard Cells experiment heat is conducted before the convective rolls emerge.

\item As the gradient imposed by the potential energy of the water is increased beyond the first critical value $\rho=1$, a supercritical Pitchfork bifurcation occurs. The fixed point loses its stability and the wheel starts to accelerate until it reaches a fixed limit angular velocity. This value corresponds to $\omega^{*}=\pm K\sqrt{\rho-1}$, $a_1^{*}=\pm q_1\sqrt{\rho-1}/K\rho$ and $b_1^{*}=q_1/K\rho$. The system is locally active now, maintaining a steady coherent motion far from equilibrium. Two facts deserve attention. Firstly, the rotational velocity at equilibrium increases following a square root trend as we drift apart from the unstable state $\omega^{*}=0$, by increasing $\rho$. Secondly, the choice of the sense of rotation is, though deterministic, for all practical purposes, uncontrollable. Any fluctuation at the microscopic scale can decide which sense of rotation is accomplished as the bifurcation is traversed. Manifestly, the critical bifurcation induces a symmetry breaking phenomenon concerning the group of the parity transformation of the wheel. Once an angular velocity settles, it remains stable under perturbations smaller in size than such value of the speed (see Fig.~\ref{fig:2} again). This stationary rotation portrays homeostasis because the system reacts to small external perturbations through negative feedback  \cite{ste08}. This reaction prevents a switch between the two arms illustrated in the bifurcation diagram.

\item Stability analysis of Eqs.~\eqref{eq:5} can show that, as the system is progressively driven beyond the critical point $\rho_c=\sigma(\sigma+4)/(\sigma-2)$, a subcritical Hopf bifurcation takes place. This second bifurcation destabilizes the two stable arms yielding a preexisting chaotic motion, which encompasses both equilibrium points, as the only possible asymptotic limit set. This strange attractor is the famous Lorenz attractor, depicted in Fig.~\ref{fig:3}. It is stable as a whole so that any perturbations driving the system away are dissipated as time goes by. However, now the symmetry of the system is broken everywhere since all the trajectories inside the attractor are unstable. Little perturbations eventually make nearby trajectories diverge. As shown ahead, on average, the thermodynamic behavior of the system in this third domain, which is posed very far from equilibrium, is not truly different from what we observed in region two, except for the enhanced sensitivity to external fluctuations. This internal instability makes the system more adaptable to changes in the environment.
\begin{figure}
\centering
\includegraphics[width=0.7\linewidth,height=0.6\linewidth]{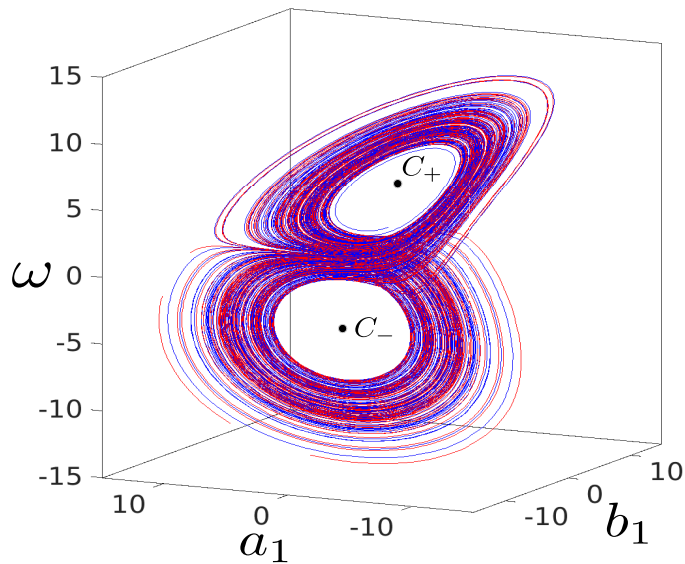}
\caption{\textbf{Lorenz attractor}. A strange attractor representing the chaotic spinning of the wheel, unpredictably switching from clockwise $C_{+}$ to anticlockwise $C_{-}$ rotational speeds.}
\label{fig:3}
\end{figure}

\end{enumerate}

\section{Thermodynamic analysis}

We now proceed to derive analytical expressions for some thermodynamic properties of the waterwheel. In particular, we shall focus on three fundamental quantities, which have been addressed in previous empirical studies with B\'enard Cells \cite{sil57}: the thermodynamic efficiency, the entropy generation, and the exergy destruction. We can comparatively study these properties in the waterwheel, which is more faithfully represented by the Lorenz equations, relating them to the empirical results of the more difficult spatiotemporal problem of fluid convection. This relation will entail a discussion about some theoretical conjectures that have been exposed in previous works \cite{sch94}, concerning a proposal of a restated second law of thermodynamics for open systems. Mainly, as such systems are pushed far away from equilibrium by some external gradients, they take advantage of all available means to resist or to compensate them. Therefore, these authors propose that as ecosystems grow and develop, they tend to increase their total dissipation to abet the highest exergy degradation, self-organizing themselves into progressively complex dissipative structures, with increased energy flow and cycling activity. Arguably, these structures can later couple among them, generating greater diversity at the new hierarchical levels.
\begin{figure}
\centering
\includegraphics[width=0.7\linewidth,height=0.3\linewidth]{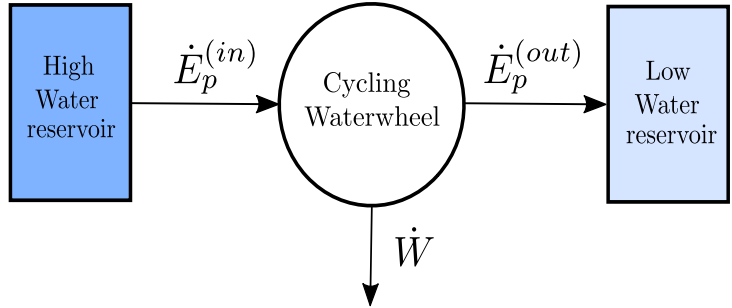}
\caption{\textbf{The wheel as an engine}. The Malkus-Lorenz wheel as a thermodynamic machine that is fed by water with high potential power from a reservoir (the hose) and that delivers water to another reservoir with low potential energy at some rate. As in traditional heat engines, the rate of power generation can be related to the difference between incoming and outgoing energy rates.}
\label{fig:4}
\end{figure}

\subsection{Thermodynamic efficiency}

The computation of the thermodynamic efficiency of the waterwheel is achieved as follows. As depicted in Fig.~\ref{fig:4}, the system ingests water with high potential energy, and with its motion takes it to the bottom, where it can be evacuated at lower gravitational energy. Thus we have a thermodynamic machine operating between two reservoirs of matter, the hose at the top and the pipes enabling the leakage of water. Since the temperature and the pressure of the water barely change with the rotation of the wheel, we can neglect any changes in the specific internal energy of the fluid. Moreover, once the motion has converged to its asymptotic limit set, the internal energy of the water remains constant, so that $\dot{U}_s=0$. If the dynamic of the wheel is chaotic, we can still consider that this equality stands because, as can be shown using Eqs.~\eqref{eq:5}, the relation $\dot{m}_{in}=-\dot{m}_{out}$ holds. Then, according to the first law of thermodynamics, we can write the energy balance of the wheel as
\begin{equation}
\dot{W}=\dot{E}_p^{(in)}-\dot{E}_p^{(out)},
     \label{eq:7}
\end{equation}
where the minus sign has been introduced to account for the fact that the power assumed from the hose is absorbed, while the power through its leakage is emitted. As explained below, we have neglected the loss of heat produced by the magnetic brake because it compensates with the work done by its friction. Therefore, the efficiency $\eta(t)$ of the waterwheel is given by the ratio of the power delivered to the rotation and the energy taken from the reservoir $\eta=\dot{W}/\dot{E}_{p}^{(in)}$. This computation yields the equation
\begin{equation}
\eta=1-\dfrac{\dot{E}_p^{(out)}}{\dot{E}_p^{(in)}}.
     \label{eq:8}
\end{equation}
Later on, the power delivered to the rotation is converted to heat through the dissipation of heat produced by the magnetic breaks. This effect produces a uniform motion that, otherwise, would accelerate without bound. In this case, the efficiency is constant, while in the case of chaotic dynamics, we have to study the average efficiency along the entire strange attractor. We stress that thermodynamic efficiency is a dynamical variable. 

Thus we have to compute the potential power injected to and released by the Malkus-Lorenz wheel. This can be done as follows with some effort. The mass of water at the top enters at a rate $Q(\theta)$, thus the potential energy of the incoming mass comprised by two infinitesimally close points of the hose is $d E_{p}^{(in)}=Q(\theta) g h(\theta)d\theta$, where $h(\theta)=r\sin\alpha(1+\cos\theta)$ is the height of the compartment at the angular position $\theta$. Therefore, we have that the rate of energy inflow is
\begin{equation}
\dot{E}_p^{(in)}=g \int^{2 \pi}_{0}Q(\theta)h(\theta) d\theta.
     \label{eq:9}
\end{equation}
For simplicity, we assume a symmetric arrangement for the hose, with $Q(\theta)=q$ in the range $|\theta|<\beta$. This yields
\begin{equation}
\dot{E}_p^{(in)}=2 g r \sin \alpha q (\beta+\sin \beta).
     \label{eq:10}
\end{equation}

On the other hand, the mass of water is leaked uniformly at a fractional rate $K$, through any point placed along the ring of the wheel. Since the mass of water in the compartment spanning the interval ($\theta$,$\theta + d\theta$) is $d m=\mu(\theta) d\theta$, this computation is even simpler, leading to 
\begin{equation}
\dot{E}_p^{(out)}=g K\int^{2 \pi}_{0}\mu(\theta)h(\theta) d\theta.
     \label{eq:11}
\end{equation}
Only the two first even coefficients of the Fourier expansion of the mass density contribute to the expel of power in terms of mass of water. The results is 
\begin{equation}
\dot{E}_p^{(out)}=\pi g r \sin\alpha K (2 b_0+b_1).
     \label{eq:12}
\end{equation}
Therefore, we get the thermodynamic efficiency
\begin{equation}
\eta=1-\dfrac{\pi K (2 b_0+b_1)}{2 q (\beta+\sin \beta)}.
     \label{eq:13}
\end{equation}
In the second region shown in the bifurcation diagram, where the rotation of the wheel is uniform, we can solve Eq.~\eqref{eq:4} and, using the two first coefficients of $Q(\theta)$, which are $q_0=\beta q /\pi$ and $q_1=2 q \sin\beta /\pi$, obtain the fixed value of the efficiency in a closed form
\begin{equation}
\eta^{*}=\dfrac{\sin \beta}{\beta + \sin\beta}\left( 1-\dfrac{1}{\rho} \right).
     \label{eq:14}
\end{equation}

In the limit in which all the water is poured through a single hole $\beta \rightarrow 0$, we have $\eta^{*}=(1-1/\rho)/2$. This equation tells us that the efficiency of the wheel as the system is pushed far from equilibrium is monotonically increasing. Note that the maximum efficiency achievable corresponds to the limit $\rho \rightarrow \infty$, which yields an upper bound of $\eta_{max}=1/2$. In the forthcoming sections, we relate this value to the Carnot's efficiency of the machine \cite{car72}. Despite the appearances, this value is not too bad for a motor simply driven by a flow of water. Typical values of efficiencies of heat engines range between $25\%$ and $50\%$ \cite{cen11}. The second-law efficiency $\eta_{II}$ is frequently defined to compute the efficiency of a thermodynamic engine relative to its optimum level of operation $\eta/\eta_{max}$, which corresponds to reversible processes \cite{cen11}. This concept is relevant because it is related to the exergy recovered by the machine \cite{cen11}. In our case, this renormalized concept of thermodynamic efficiency yields the marvelous expression
\begin{equation}
\eta^{*}_{II}= 1-\dfrac{1}{\rho},
     \label{eq:15}
\end{equation}
a formula that evokes the efficiency of the Carnot's cycle, where the quotient of temperatures between the heat and the cold reservoirs has been simply replaced by the quotient of income and outcome rates of water flow.

In the case of chaotic dynamics we can not derive a closed expression for the efficiency because it evolves through the chaotic attractor. Nevertheless, we can average the efficiency as the limit set is densely covered by a particular trajectory. This entails us to write the average thermodynamic efficiency as
\begin{equation}
\langle \eta \rangle=\dfrac{\sin \beta}{\beta + \sin\beta}\left( 1-\langle b_1 \rangle \dfrac{K}{q_1} \right),
     \label{eq:16}
\end{equation}
where the average of a function $f(t)$ is computed for non-periodic trajectories as follows
\begin{equation}
\langle f \rangle=\lim_{\tau \rightarrow \infty}\dfrac{1}{\tau} \int_{0}^{\tau}f(t)dt,
     \label{eq:17}
\end{equation}
always remaining bounded for strange attractors.
\begin{figure}
\centering
\includegraphics[width=0.4\linewidth,height=0.2\linewidth]{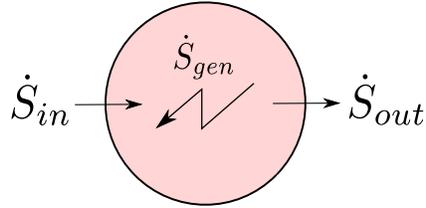}
\caption{\textbf{Entropy analysis}. In the present figure we schematically represent the entropy analyis of a dissipative structure according to the second law of thermodynamics. The structure absorbs entropy through the income of heat and matter, and absorbs negentropy by the outcome of heat and matter through some other channels. The total rate of entropy generation is $\dot{S}_{s}=\dot{S}_{in}+\dot{S}_{out}+\dot{S}_{gen}$. For stationary homeostatic systems $\dot{S}_s=0$, so that the generation of entropy balances with the entropy interchanged with the environment.}
\label{fig:5}
\end{figure}

\subsection{Entropy generation}

A crucial thermodynamic variable in the study of dissipative structures is the rate of entropy generation $\dot{S}_{gen}$. This magnitude measures the degree of irreversibility of the system. As it has been recently outlined \cite{lop20,lop21}, this fundamental irreversible character of open physical systems might have its origin in the irretrievable radiation dissipated to the ambient that, as we know thanks to the black body experiment, is tantamount to heat \cite{bon17}. Furthermore, entropy generation is intimately related to exergy destruction by a thermodynamic system. Exergy $X$ is the maximum available work (hence reversibly obtained) that can be extracted from a system in disequilibrium with its environment \cite{cen11,bon17}. Correspondingly, whenever a physical system presents non-conservative forces, this capacity to do work is limited in some way or another due to losses. Because external gradients are imposed on dissipative structures, exergy, as a thermodynamic potential to perform work, has to be of the uttermost relevance in open physical systems. This contrasts with closed systems in thermodynamic equilibrium with their environment (\emph{e.g.} Helmoltz's free energy $F$ is well suited for closed physical systems in thermal equilibrium, while the Gibbs' potential $G$ is mandatory when studying chemical equilibrium).

In general terms, the second law for open systems must be written in the form
\begin{equation}
\dot{S}_s=\dot{S}_e+\dot{S}_{gen},
     \label{eq:18}
\end{equation}
with $\dot{S}_s$ the rate of rate of entropy variation in the system, $\dot{S}_e$ the entropy interchange with the surroundings, and $\dot{S}_{gen}$ the entropy production rate \cite{pri67}. As it is well-known, the second principle enforces the inequality $\dot{S}_{gen} \geq 0$. However, a system can decrease or increase its entropy by interchanging it with its environment. In general terms, dissipative structures operate at some stable stationary point, so that they do no change their entropy $\dot{S}_{s}=0$, what again can be identified with the physiological concept of homeostasis \cite{ste08}. This implies that \emph{negentropy} must be attained $\dot{S}_{e} \leq 0$ by a release of entropy to the environment in terms of an outward flow of matter or through transpiration by heat. Even if the dynamics is chaotic at some point, on the average we can consider $\langle \dot{S}_{s} \rangle =0$, as shown ahead. Thus the present analysis can be easily extended to the chaotic realm.
 
In general, the rate of entropy interchange with the environment includes all incoming and outgoing entropy flows in the form of matter and heat. In the present case, there exists a flow of entropy into the system in terms of a mass of water. The rate of entropy income is $\dot{S}_{in}=\dot{m}_{in} s$, where $s$ is the specific entropy of the water. Since this specific entropy barely changes during the motion of the wheel, we can approximate the rate of entropy ejection due to matter and heat as $\dot{S}_{out}=\dot{m}_{out} s+\dot{Q}/T$, where $\dot{Q}$ is the heat dissipated to the environment in terms of electromagnetic radiation coming from the Foucault currents inside the magnetic brake and $T$ is the temperature of the environment. In the stationary state $\dot{m}_{in}=-\dot{m}_{out}$, thus we have that the matter contributions cancel each other. According to the first principle of thermodynamics or, more precisely, to the equivalence between heat and mechanical work, $\dot{Q}=-\dot{W}_{fric}$, what implies that $\dot{S}_e=-\nu \omega^2/T$, where we recall that $\nu$ describes the amount of friction. We assume that the wheel is in equilibrium with the surroundings, neglecting the temperature gradient in the immediate air around the brake. Therefore, following Eq.~\eqref{eq:18}, in the stationary state far from equilibrium, we obtain
\begin{equation}
\dot{S}_{gen}=\frac{\nu \omega^2}{T}.
     \label{eq:19}
\end{equation}
Furthermore, in the region of the parameter space where the rotational motion is periodic, we simply have
  \begin{equation}
\dot{S}_{gen}^{*}=\frac{\nu K^2}{T}(\rho-1),
     \label{eq:20}
\end{equation}
as long as the Rayleigh number $\rho$ is greater than one and smaller than the critical value $\rho_c$, before the onset of chaos. Therefore, the generation of entropy per cycle can be written as
  \begin{equation}
\Delta S^{*}_{gen}=\frac{2 \pi\nu K}{T}\sqrt{\rho-1}.
     \label{eq:21}
\end{equation}
It is evident that $\Delta S^{*}_{gen}$ tends to zero as $\nu$ goes to zero, as expected in the case of a reversible machine. However, we notice that $\dot{S}_{gen}$ does not go to zero, which defies our intuition. This conflict lies in the following fact. As the friction tends to vanish, the rotational speed grows unbounded, and the machine's cycle occurs instantaneously. Here we see a significant difference with the reversibility frequently imposed in equilibrium thermodynamics by making infinite-lasting cycles that go through a series of very small steps, during which the system interchanges energy with a sequence of external reservoirs. On the contrary, when studying far from equilibrium phenomena, cyclic processes last a finite amount of time. For this to happen, dissipation is essential to avoid the divergence of the angular velocity to infinity as time goes by, according to Newton's second law, which has been recently derived from Maxwell's classical electrodynamics \cite{lop20,lop21}.
 
More importantly, we highlight that, as the system is progressively driven far from equilibrium by increasing the water inflow, the rate of entropy generation tends to increase, as it happened with the thermodynamic efficiency. This paradox is circumvented if we recall that the energy flow is increasing so that more energy can be converted to work. What is surprising is that the power of conversion in comparison to the incoming power of energy is proportionally greater as we drive the system away from equilibrium. This is a remarkable fact that constitutes the first confirmation to the claims provided in previous works \cite{sch94}, which are also in accordance with data gathered from B\'enard Cells' experiments \cite{sil57}. From a mathematical point of view, these two statements can be written as
\begin{equation}
\frac{d\eta^{*}}{d\rho} \geq 0
\end{equation}
and as $d \dot{S}_{gen}^{*}/d\rho \geq 0$, which holds everywhere except at the critical bifurcation points. A more thorough analysis of this idea is provided in the forthcoming section.
\begin{figure}
\centering
\includegraphics[width=0.61\linewidth,height=0.51\linewidth]{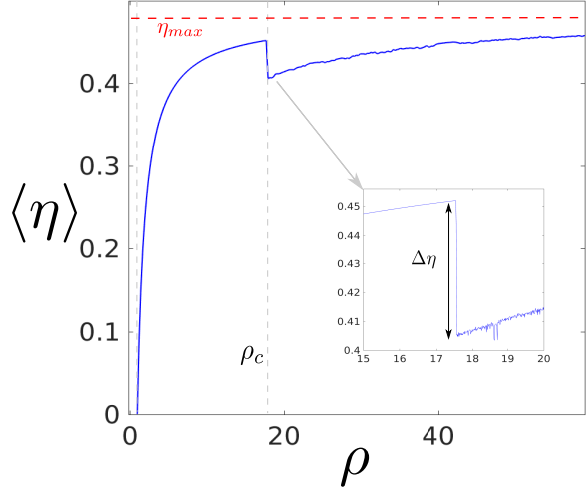}
\caption{\textbf{Thermodynamic efficency}. The value of the average efficiency as the Rayleigh number is modified by increasing the rate of water income $\rho=q_1$. As the system is pushed increasingly far from equilibrium, the efficiency increases. However, at the Hopf bifurcation $\rho_c=17.5$ a sudden drop $\Delta \eta$ is observed. The maximum value of the efficiency corresponds to $\eta_{max}=0.48$, in conformity with Eq.~\eqref{eq:14}.}
\label{fig:6}
\end{figure}

When chaotic dynamics generates the power of the wheel, the entropy generation must be averaged along the trajectory $\langle \dot{S}_{gen} \rangle=\nu \langle \omega^2 \rangle/T$, following the same recipe as before. Finally, the rate of exergy destruction can be computed straightforwardly \cite{cen11} by means of the equation $\dot{X}_{des}=T \dot{S}_{gen}$. This value constitutes the rate at which the thermodynamic cycling system is losing the opportunity to perform work through irreversibilities. Certainly, exergy destruction is a hallmark of the irreversible dynamics of dissipative structures, a price that they must pay to ensure the maintenance of their energetic and structural stability.

\section{Numerical results}

Since we want to delve deeper into the effects of chaos on the efficiency of the waterwheel, we now present results from systematic numerical simulations. For this purpose, we shall study in more detail the properties of thermodynamic efficiency and entropy generation (or exergy destruction), paying special attention to both Rayleigh and Prandtl numbers. More specifically, we compute bifurcation diagrams and explore the parameter space of the wheel by modifying the input of the water inflow $q_1$ and the degree of friction $\nu$. In order to keep a clear meaning of the parameters, we avoid non dimensionalizing them. Instead, we set the values of the remaining parameters as follows. Since the parameter $K$ always appears multiplying the value $\nu$, we consider for simplicity $K=1~s^{-1}$ hereafter. The moment of inertia can be taken as $I=1~kg\cdot m^{2}$ since this parameter can be reabsorbed by redefining the parameters $\nu$ and $r$. Following the same argument as before, we can set a value for the radius of $r= 1~m$. Again, without loss of generality, we also take an effective value of the gravitational acceleration $g_{eff}=g \sin\alpha=10/\pi~m \cdot s^{-2}$, for reasons explained below. Finally, the parameter $\beta=0.7~rad$ unless otherwise specified and the temperature of the laboratory is set to $T=300~K$. Other parameter values can be chosen if desired, yielding similar results. This parameter choice leaves us two independent fundamental parameters to study the dynamics of the wheel, which are $q_1$ and $\nu$. Using the parameter values introduced in the present paragraph, we can identify these two parameters with $\rho=10~q_1/\nu$ and $\sigma=\nu$, respectively.  
 
\subsection{Thermodynamic efficiency}

In the first place, we present the numerical computation of the thermodynamic efficiency using $\nu=10~s^{-1}$, leaving the Rayleigh number $\rho=q_1$. As shown in Fig.~\ref{fig:6}, we can clearly distinguish three regimes. When the wheel is at rest, there is no cyclic machine operating, thus the efficiency is zero. Then, from $\rho=1$, up to the critical value $\rho_c=17.5$, the machine asymptotically cycles at a constant rate, clockwise or anticlockwise. There is a monotonic relation between the rate of water inflow $q_1$ and the efficiency $\eta$ so that the further that it is pushed the system from equilibrium, the more efficient that it converts energy into work. This counterintuitive notion is related to the growth of the amplitude of the attractor's value $\omega^{*}$, which scales following a power law. The trend follows exactly the analytical result appearing in our previous study of the bifurcation diagram. 

This is not completely novel, since a similar phenomenon can be observed in traditional heat engines operating between two heat reservoirs ($e. g.$ Carnot's engine). When the temperature of the hot (cold) reservoir is enhanced (reduced), the efficiency increases. The same holds here since $\rho$ is directly proportional to $q_1$ and inversely proportional to $K$. However, since the present motor is time-irreversible, we can also tune the efficiency by changing the parameter related to its dissipative losses. Another distinction is that the variation of the parameters can trigger higher instabilities through bifurcation phenomena. At this point, the analytic equation might not hold anymore.

When sustained chaotic motion appears in the system, an unexpected dip in thermodynamic efficiency can be observed (see again Fig.~\ref{fig:6}). From a physical point of view, this occurs because the average value of $b_1$ is increased. The water is more evenly distributed and there is more water falling from the higher compartments of the waterwheel. This is correlated to an abrupt jump in the average value of $\omega^2$ due to the crisis, produced by the Hopf bifurcation. Indeed, during the turn backs of the wheel it stops and the thermodynamic efficiency reduces. In absolute terms, this corresponds to a measure of the average value of the strange attractor along such a dimension of the phase space, and, as the reader can confirm in the forthcoming section, to the rate of entropy generation. 

This consideration introduces the first nuance concerning the claims made by Schneider and Kay \cite{sch94}. Here, in contrast, we show that, despite the general tendency of the system to increase its ability to reduce the external gradient imposed on the system, the efficiency can experience dramatic changes at critical bifurcation points. It is not clear if this occurs as well in experiments about convection \cite{sil57}. This important issue deserves further exploration in many other types of dissipative structures.
\begin{figure}[H]
\centering
\includegraphics[width=0.65\linewidth,height=0.55\linewidth]{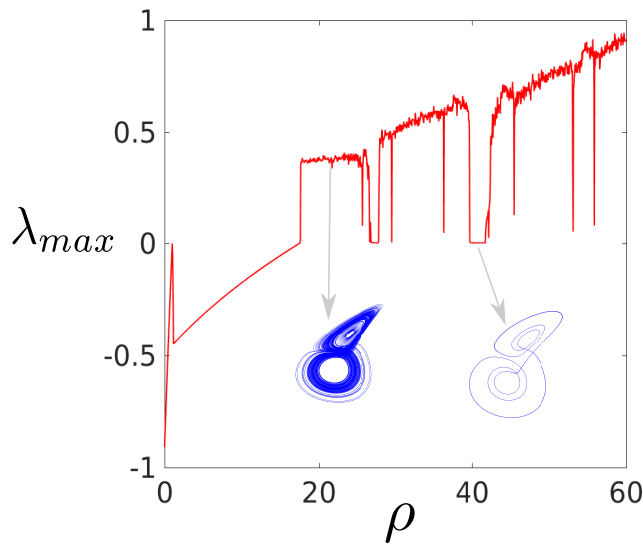}
\caption{\textbf{Largest Lyapunov exponent}. The transition to the chaotic regime occurs for $\rho>\rho_c$, when the largest Lyapunov exponent grows above zero. Importantly, we detect periodic windows with limit cycles beyond $\rho_c$, what shows that, apart from the first subcritical Hopf bifurcation, the chaotic nature of the motion is not really relevant for the thermodynamic efficiency.}
\label{fig:7}
\end{figure}

However, as the average value of the rotational speed keeps growing in size, we again see a renewed increase in thermodynamic efficiency. What we find outstanding is that the efficiency can grow even over the highest values attained in the regular regime. This fact suggests that chaos (or ``freedom", so to speak) is not necessarily harmful to thermodynamic efficiency. This idea is reinforced if we compute the value of the largest Lyapunov exponent (see Fig.~\ref{fig:7}), which displays sudden changes in windows of periodic behavior, without altering the efficiency. As long as the attractor's average value is not seriously jeopardized, as it occurs through the critical bifurcations, chaotic dynamics should not generally be an obstacle to the power production of an engine. 

\begin{figure}
\centering
\includegraphics[width=0.85\linewidth,height=0.45\linewidth]{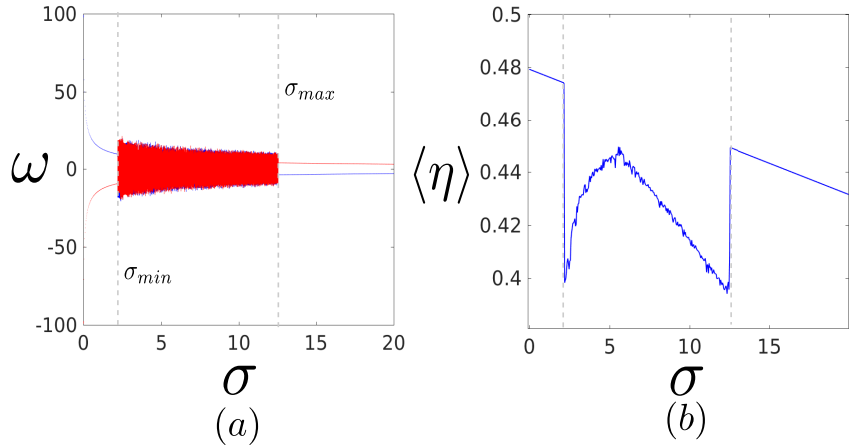}
\caption{\textbf{The effect of dissipation}. (a) We show the bifurcation diagram as the Prandtl number $\sigma=\nu/I K$ is varied by changing the value of $\nu$. A value of $q_1=20~kg \cdot s^{-1}$ has been considered. We see a transition to chaotic regime. Note how the speed of the wheel diverges to infinity as $\sigma \rightarrow 0$. The minimum and maximum values $\sigma_{min}$ and $\sigma_{max}$ delimiting chaotic dynamics can be computed from the definition of the critical value $\rho_c$. (b) The average efficiency of the wheel as the Prandlt number increases is plotted. The transitions to chaotic regime are clearly identified. At $\sigma \rightarrow 0$ the waterwheel reaches its maximum efficiency, corresponding to Carnot's value.}
\label{fig:8}
\end{figure}

For completeness, we have also computed the value of the efficiency as the parameter $\nu$ is varied. In Fig.~(\ref{fig:8}b) we show the efficiency as $\nu$ decreases from the value $\nu=20~s^{-1}$ to zero, using a value for the inflow of water of $q_1=20~km \cdot s^{-1}$. We appreciate a transition to the chaotic regime followed by a return to regular dynamics. Again, two sudden jumps can be detected as we enter and exit from the chaotic region. 

\begin{figure}
\centering
\includegraphics[width=0.89\linewidth,height=0.42\linewidth]{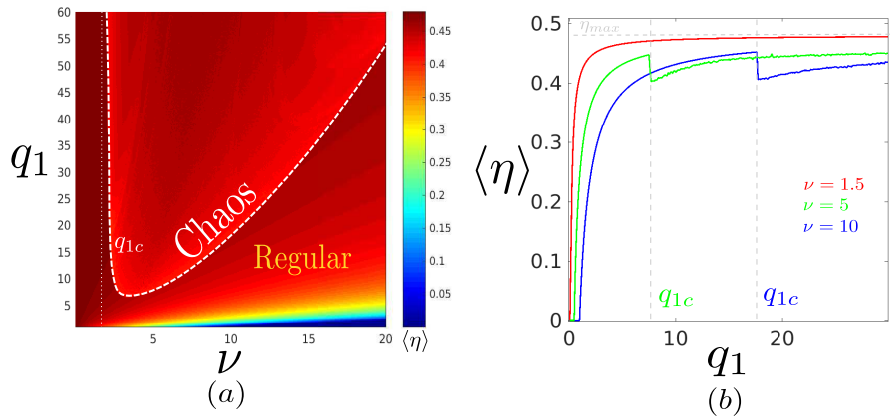}
\caption{\textbf{Efficiency parameter set}. (a) The value of the thermodynamic efficiency as a function of the inflow of water and the dissipation of the wheel has been represented in the parameter space. The dashed curve represents the transition from regular regime to chaotic dynamics and presents a vertical asymptote at $\nu=2$ (dotted line). If we delve deeper into $q_1$, periodic windows can be found, which do not affect the efficiency noticeably. (b) The efficiency for three different values of the friction $\nu$ in $s^{-1}$. As we can see, below the value $\nu=2$ there is no transition to chaotic regime and the discontinuity in the efficiency disappears (red curve).}
\label{fig:9}
\end{figure}

Interestingly, when $\nu \rightarrow 0$ we observe in Fig.~(\ref{fig:8}a) that the speed of the wheel diverges to infinity, as previously discussed. This value corresponds to a ``reversible" machine, where the maximum efficiency is achieved. Consequently, we relate this limit to the Carnot's regime of the waterwheel. The quotation marks are deserved because, even though $\nu$ makes the waterwheel reversible from a thermodynamic point of view, the dynamical system is still dissipative, insofar as the trace of the jacobian of the system of Eq.~\eqref{eq:5}, which is equal to $-(2K+\nu)$, is negative as long as $K>0$. In other words, there are two contributions to the irreversibility of the dissipative structure from a dynamical point of view: the loss of heat and the loss of water. Thus, in general, we see that it is irretrievable losses that shoot the arrow of time of dissipative structures \cite{lop20,lop21}. However, this fact does not necessarily imply that open dynamical systems tend to dead states, since energy can be fed into them as well. In Fig.~(\ref{fig:9}a) the efficiency in the parameter space is presented, where a sharp transition defined by the Rayleigh's critical value $\rho_c$ as a function of $\sigma$ is insinuated. This critical curve separates the region of parameter space where regular regime occurs, from the region where chaotic dynamic can take place. Importantly, we see that this curve has a vertical asymptote for $\nu=2~s^{-1}$. Below this value the system does not transit to chaotic regime as $q_1$ is increased and there is no phase transition, as shown in Fig.~(\ref{fig:9}b). Therefore, the efficiency curves in this regime do not present any discontinuous jump as the water inflow is increased and the efficiency monotonically raises towards its maximum output.

\subsection{Entropy generation}

To conclude our thermodynamic analysis, we corroborate that the critical drop in the efficiency is explained by the trend of the square rotational speed $\omega^2$ of the waterwheel, which is proportional to the rate of entropy generation. This computation will also bear information about the trends of entropy generation in the chaotic regime that, on average, does not change when we compare it to the regular one. 

\begin{figure}
\centering
\includegraphics[width=0.6\linewidth,height=0.5\linewidth]{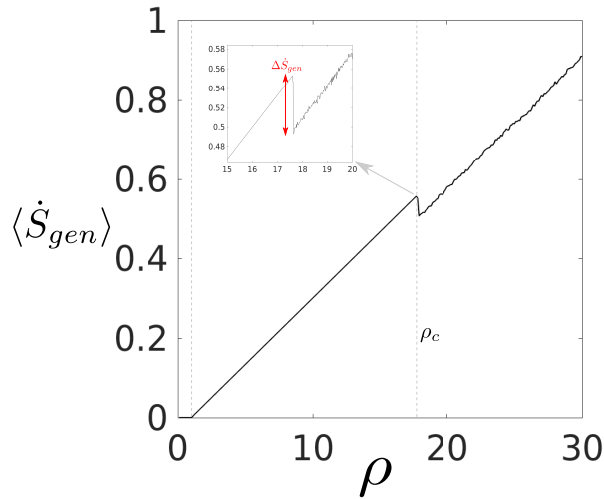}
\caption{\textbf{Entropy generation}. The average rate of entropy generation of the system as it is pushed far from equilibrium tends to linearly increase. A slump is observed when the system transits to chaotic regime, correlating entropy generation and thermodynamic efficiency.}
\label{fig:10}
\end{figure}

In Fig.~\ref{fig:10} we depict the average rate of entropy generation versus the key parameter $\rho$. As previously demonstrated through mathematical analysis, the tendency in the regular regime is linear. We also note how this tendency is preserved on average during the chaotic regime. And again, we also find an abrupt change when the Hopf bifurcation takes place. In this manner, there is an interesting correlation between thermodynamic efficiency and the rate of entropy generation. 

Bearing in mind that we are considering linear phenomenological laws, it is expected that as generalized fluxes transit through critical bifurcation points in non-equilibrium thermodynamic systems, discontinuous jumps in the entropy generation of the system shall arise. This phenomenon is reminiscent of other first or second order phase transitions commonly appearing in classical thermodynamics. For example, the self-organization of a fluid into convective cells gives rise to the appearance of a characteristic length within a periodic structure, what doubtlessly resembles the formation of a crystal. Later on, as the temperature gradient increases, the convection cells become turbulent and this crystal melts, so to speak. Or, to bear another example of historical relevance, the paramagnetic-ferromagnetic phase transition frequently represented by the Ising model \cite{ons44} can be regarded as a Pitchfork bifurcation in the order parameter. Simply put, we believe that phase transitions should generally be regarded as dynamic bifurcations.

Apart from the aformentioned jump, the fact that the tendency is linear opposes to experimental results made with B\'{e}nard Cells. This conflicts with a claim made in previous works \cite{sch94}. Namely, that exergy destruction, when plotted against the adimensional parameters that drive the system out of equilibrium, should have a positive curvature. From a physical point of view, this is interpreted by these authors in the following way. As a system is pushed further from equilibrium, the efforts made by the system to compensate the external gradients are greater in proportion. Our dynamical system constitutes a counterexample to this statement, since the equation $d^2 \dot{S}_{gen}/d\rho^2 >0$ does not hold. In general, we expect that the curvature of the rate of entropy generation curve should be related to the curvature of the branches appearing in the bifurcation diagrams. This is because linear phenomenological relations (together with Onsager's reciprocal relations, if desired) express the rate of entropy generation as a function of the square of the generalized fluxes \cite{pri67}. Then, for such a statement to be true, the bifurcation branches of the generalized fluxes should grow as a power higher than one-half ($e.g.$, the angular velocity $\omega$ as a function of $\rho$, in the case case at study). Many other counterexamples are expected within the theory of dynamical systems \cite{ali96}, and, in principle, we cannot elucidate a mathematically compelling reason why values even smaller than $1/2$ should be disallowed, depending on the particular nature of the bifurcation branches that far from equilibrium. Perhaps, this departure between the Lorenz model and the experimental data gathered from studies of hydrodynamic convection indicates that nonlinear phenomenological laws (\emph{e.g.} a quadratic term of friction) must be considered when the system is far enough from thermodynamic equilibrium.

\section{Discussion}

In the present work we have shown that the proposal of a generalized second law for open thermodynamic systems expressed in a positive way, and not just as a law stating the inaccessibility of certain states, is plainly justified if we consider as central the concepts of thermodynamic efficiency and exergy destruction \cite{sch94}. Indeed, we have provided analytical and nummerical evidence that, as our dynamical system is driven far from equilibrium, the thermodynamic efficiency tends to increase. Furthermore, we have also confirmed a positive correlation between thermodynamic efficiency and entropy production or, equivalently, exergy destruction \cite{sch94}. However, our study introduces several particular amendments to some statements claimed in this essay. Firstly, the efficiency can experience abrupt changes at critical bifurcation points. Secondly, and in accordance with the previous fact, the rate of entropy generation does not increase at such transitions, neither it always increases with positive curvature, even if it does certainly increase elsewhere. Finally, we have also shown that the presence of chaos does not reduce the thermodynamic efficiency, as long as the average value of the generalized fluxes over the limit cycle attractor does not decrease.

It was originally proposed by Boltzmann and later on by Lotka \cite{lot22a,lot22b} that natural selection gives an advantage to those organisms whose energy-capturing devices are most efficient seizing external available energy and funneling it into channels that favor the survival of the species. This concept was later reconsidered by Odum, who coined the expression \emph{maximum power principle} to denote this statement \cite{odu55}. According to this author, the thermodynamic efficiency and the total power produced by a physical system should be both considered the physical magnitudes upon which the environment exerts a selective pressure, favoring those structures that increase overall energy flow \cite{odu63}. Here, we have related this principle to a very specific mathematical instance: the second-law thermodynamic efficiency. Apart from the size of the system $r$, this concept incorporates the rate inflow of matter $q$, the rate of energy delivered to the environment $K$, and the degree of dissipation $\nu$. The former increase the entropy of the thermodynamic structure, while the latter two produce negentropy. In homeostatic structures, the last of the three is also related to the entropy generation in the thermodynamic system. The three elements can be tuned by a dissipative structure to better accommodate and synchronize to other elements in its environment. 

To conclude, and following Lotka's principle, we propose that Darwinian evolution favors those dissipative structures that are thermodynamically more efficient. We also draw attention to the fact that such structures usually arrange integrating large complex networks, which are coupled through diffusive currents. These wirings entail the formation of complex patterns and the propagation of nonlinear waves through the ecosystem if the parameters of these structures are tuned to values that are in the locally active domain of the parameter space \cite{mai13}. The study of the effect of such connectivity and the resulting dynamics on the overall thermodynamic efficiency is of paramount importance. As larger chains are attained, more work-producing steps are added, allowing to build up larger and more efficient \emph{exergy cycles} capable of reducing the energy gradient imposed by the sun \cite{sch94}. We wonder the extent to which profound stresses exerted on these strongly coupled cycles could damage the thermodynamic efficiency of the biosphere \cite{sch94}.

\section{ACKNOWLEDGMENTS}
The authors wish to thank \'{A}lvar Daza and Rub\'{e}n Capeans for fruitful discussion on Carnot's theorem and the concept of reversible processes.

\begin{appendices}
\renewcommand{\theequation}{A\arabic{equation}}
\setcounter{equation}{0} 

\end{appendices}

\end{document}